\documentclass[final,5p,times,twocolumn]{elsarticle}

\usepackage{amsmath}
\usepackage{amssymb}
\usepackage{graphicx}
\usepackage{color}
\usepackage{hyperref}

\usepackage[mathscr,scaled=1.15]{urwchancal}
\DeclareFontFamily{OT1}{pzc}{}
\DeclareFontShape{OT1}{pzc}{m}{it}%
{<-> s * [1.15] pzcmi7t}{}
\DeclareMathAlphabet{\mathpzc}{OT1}{pzc}{m}{it}

\definecolor{purple}{rgb}{0.5,0,0.5}
\definecolor{blue}{rgb}{0.0,0,0.9}

\biboptions{sort&compress}

\journal{Physics Letters B}

\begin{document}

\begin{frontmatter}

\title{Pion Valence-quark Parton Distribution Function}

\author[UA]{Lei Chang}
\author[UA]{Anthony W. Thomas}

\address[UA]{CSSM, School of Chemistry and Physics
University of Adelaide, Adelaide SA 5005, Australia}
\date{30 Oct 2014}

\begin{abstract}
$\,$\\[-7ex]\hspace*{\fill}{\emph{Preprint no}. ADP-14-30/T889}\\[1ex]
Within the Dyson-Schwinger equation formulation of QCD, a rainbow ladder 
truncation is used to calculate the pion 
valence-quark distribution function(PDF). 
The gap equation is renormalized at a typical hadronic 
scale, of order 0.5GeV, 
which is also set as the default initial scale for the pion PDF. 
We implement a corrected leading-order expression for the PDF which 
ensures that the valence-quarks carry all of the pion's light-front momentum 
at the initial scale. The scaling behavior of
the pion PDF at a typical partonic scale of order 5.2GeV 
is found to be $(1-x)^{\nu}$, 
with $\nu\simeq 1.6$, as $x$ approaches one. 
\end{abstract}


\end{frontmatter}
\medskip

Given its dual roles as a conventional bound-state in quantum field theory 
and as the Goldstone mode associated with dynamical
chiral symmetry breaking, the pion has been proven critical to explaining 
phneomena as diverse as the long-range nucleon-nucleon interactions
and 
the flavor asymmetry observed in the quark sea of the 
nucleon~\cite{Thomas:1983fh}. 
The study of the pion
structure function is of great interest as a 
fundamental test of our understanding of nonperturbative QCD. 
Experimental information on the
parton distribution function(PDF) in the pion has primarily been 
inferred form the Drell-Yan reaction in 
pion-nucleon collisions\,\cite{Badier:1983, Betev:1985, Conway:1989, Wijesooriya:2005}. 

Lattice QCD calculations\,\cite{Brommel:2007, Best:1997, Detmold:2003} 
have traditionally been able to yield only the 
low-order moments of the PDFs. 
While there has been a recent suggestion of a very promising 
way\,\cite{XJi:2013} to directly compute the $x$-dependence 
in lattice QCD, it will take considerable 
effort to reliably extract the large-$x$ behavior using this method. 
The calculation of PDFs within models is challenging and various models
have given a diversity of results. 
Most models, including the QCD parton model\,\cite{Farrar:1975}, 
pQCD\,\cite{pQCD} and 
the Dyson-Schwinger equations(DSE)\,\cite{Hecht:2001PRC, Nguyen:2011PRCR} 
indicate that at high-$x$ the PDF should behave as $(1-x)^{\alpha}$, 
with $\alpha \simeq 2$. The Nambu-Jona-Lasinio(NJL) models\,\cite{NJLsum} 
with translationaly 
invariant regularization and Drell-Yan-West
relation\,\cite{DYWrelation} favors a linear dependence on $1-x$. 

The first DSE study of the pion PDF was based~\cite{Hecht:2001PRC} upon an
analysis that employed phenomenological parametrizations 
of both the Bethe-Salpeter amplitude and the dressed-quark propagators.
A numerical solution of the DSE utilizing the rainbow-ladder(RL) 
truncation has been used to compute
the pion and kaon PDFs following same line~\cite{Nguyen:2011PRCR}.

In this work we revisit the pion valence PDF within the DSE approach,  
with the following improvements: 1) the rainbow-ladder
gap equation is renormalized 
at a typical hadron scale, $\zeta_{H}$, that also 
serves as the initial scale for the PDF; 2) a corrected
leading-order expression for the PDF is employed within the RL 
trunction; 3) the extraction of the PDF is based on its moments,
a method that has been widely used in parton distribution amplitude 
calculations\,\cite{Chang:2013pq, Shi:2014Kaon, Gao:2014rho}. 
The large-$x$ behavior is naturally
reflected in the high moments. In the method used here, 
we can calculate any large moment and thus we have a 
reliable tool with which to analyze the large-$x$
behavior. 

In order to help explain the numerical results and place them 
in some perspective, we introduce several
models which produce pointwise PDFs.
Our suggestions cover a broad range of possibilities, against
which the predictions of the present model may be compared, especially 
calculations that can be described
within the amplitude language, such as the DSE and NJL models 
with various regularization frameworks.  

In  Ref.\,\cite{Chang:2014Basic} a corrected, leading-order expression 
was given for the pion's valence-quark PDF. 
This expression produces the model-independent result that quarks dressed via 
the RL truncation carry all of the pion's light-front momentum at 
a characteristic hadronic scale,
if the meson amplitude is momentum dependent. 
We quote the form of the quark distribution function 
in the RL truncation here:
\begin{eqnarray}
\label{TqFULL}
 q(x)&=&N_c  {\rm tr} \int_{dk}^{\Lambda}\!
\delta(n\cdot k-x n\cdot P)\,
\partial_{k} \left[ \Gamma(k-\frac{P}{2};-P)S(k)\right]\nonumber\\&& \Gamma(k-\frac{P}{2};P) S(k-P)\,,
\end{eqnarray}
%
In the infinite momentum frame, $q(x)$ is the number density for 
a single parton of flavor $q$ to carry the momentum 
fraction $x=n\cdot k/n\cdot P$, 
which is positive definite over the physical region $0<x<1$. 
Here, $n$ is a light-like four-vector, $n^{2}=0$; 
$P$ is the pion's four-momentum, $P^{2}=-m_{\pi}^{2}$, 
with $m_{\pi}$ the pion mass; 
$\int_{dk}^{\Lambda}$ is a Poincar\'e-invariant regularization of 
the four-dimensional momentum integral (over $k$), with $\Lambda$ the
ultraviolet regularization mass-scale. In addition, $S$ and $\Gamma$ 
are the quark propagator and pion Bethe-Salpeter amplitude,
respectively. In the present work the ultraviolet behaviour of 
$S$ and $\Gamma$ is controlled by the one-gluon exchange
interaction. In this case the above integral is ultraviolet divergence free 
and $\Lambda$ can be set to infinity safely.

As the derivative in Eq.~(\ref{TqFULL}) acts on the full 
expression within the brackets
it naturally yields two terms. The term related to the derivative of
the quark propagator yields the so-called impulse-approximation.
That corresponds to the textbook ``handbag'' contribution
to virtual Compton scatering. The second term, arising from the action 
of the derivative on the amplitude originates in
the initial/final state interactions. This expression is the 
minimal expression that retains the contribution to the quark 
distribution function 
from the gluons which bind dressed-quarks into the meson. 
This contribution may be thought of as a natural 
consequence of the nonlocal properties of the
pion wave function. That is, it expresses the process where a photon 
is absorbed by a dressed quark, which then 
proceeds to become part of the pion bound-state 
before re-emitting the photon. 
It is easy to prove that the distribution function is symmetric,
$q(x)=q(1-x)$, under isospin symmetry and the valence quarks 
carrry all of the momentum of the meson.

We describe pion as bound state in quark-antiquark scattering, using the 
Bethe-Salpeter equation. This takes the abbreviated form:
\begin{equation}
 \Gamma_{\pi}(k;P)=\int_{dq}^{\Lambda}K(q,k;P) \chi_{\pi}(q;P)
\end{equation}
where $q$ and $k$ are the relative momenta between the quark-antiquark pair, 
$P$ is the pion's four momentum and 
\begin{equation}
\chi_{\pi}(q;P)=S(q_{+})\Gamma_{\pi}(q;P)S(q_{-}) 
\end{equation}
is the pion's Poincar\'e-covariant Bethe-Salpeter wave-function, 
with $\Gamma_{\pi}$ the Bethe-Salpeter amplitude. 
Using isospin symmetry we label the dressed quark propagators $S(q_{\pm})$,
where
$q_{\pm}=q\pm \frac{P}{2}$, without loss of generality. Explicitly, these 
take the form:
\begin{equation}
 S^{-1}(k)=i\gamma\cdot k A(k^{2})+B(k^{2})
\end{equation}
where the scalar functions $A,B$ depend on both momentum and 
the choice of renormalization point.

In this work, we perform the ladder truncation for the quark-antiquark 
scattering kernel, $K(q,k;P)$.
This approximation has been widely used to
compute the spectrum of meson bound states and related properties. 
In this framework the quark-gluon vertex is bare and
a judicious choice of effective gluon propagator provides  
a connection between the infrared and ultroviolet scales.
We use the interaction provided in Ref.\,\cite{Qin:2011model}, 
which contains two different parts.
Its ultraviolet composition preserves the one-loop renormalization 
group behavior of QCD so that, 
as we shall see,
the leading Bethe-Salpeter amplitude takes 
the well known, model independent ultraviolet behavior. 
The parameters of the infrared interaction, $D\omega$ and $\omega$,  
manifest the strength and width 
of the interaction, respectively. It is chosen deliberately to be consistent 
with that determined in modern studies of the gauge sector of QCD.

The rainbow ladder truncation of the DSEs preserves the chiral symmetry 
of QCD. The renormalization
constants for the wave function and mass function $Z_{2,4}(\zeta,\Lambda)$ 
must be included to regularize the logarithmic ultraviolet
divergences. In the present calculation we follow the current quark mass 
indepedent renormalization approach
introduced in Ref.\,\cite{Remorm:massless}. In practice, 
the renormalization constant should be determined consistently 
by the condition $A(k=\zeta)=1$ 
and $\frac{\partial B(k=\zeta)}{\partial m_{\zeta}}=1$ in the chiral limit.
It should be noted that the renormalization point can be chosen 
in either the ultraviolet or infrared region and the 
quark mass function is independent of this choice. 
The renormalization constant decreases as the scale decreases,
reflecting the increase in the coupling strength in the infrared region.
Here we choose $\zeta=0.5$GeV. 

Before discussing any numerical results, it is interesting to 
recall some general features of the shape of the 
PDF and the Bethe-Salepter amplitude for a meson.
It has been shown that the significant features of $q(x)$, 
in Eq.\ref{TqFULL}, can be illustrated algebraically with some simple
models. To take a close look at the relation 
between $q(x)$ and the pion Bethe-Salpeter amplitude, we 
consider an algebraic model where the quark propagator
is contact-like and the meson amplitude expressed by its ultraviolet form 
\begin{equation}
 S^{-1}(k)=i\gamma\cdot k + M
\end{equation}
and
\begin{equation}
 \Gamma_{\pi}(k;P)=i\gamma_{5}\frac{12}{5}\frac{M}{f_{\pi}}\int_{-1}^{1}
d z \rho(z)\frac{M^{2}}{k^{2}+z k\cdot P+M^{2}} \, , 
\end{equation}
where $M$ is a dressed-quark mass, $f_\pi$ is the pion 
decay constant and we focus on the case of a massless pion. 
The factor $12/5$ is the normalization constant needed 
to ensure the charge conservation. 
$k$ is the relative momentum of the quarks in the pion and
we choose the amplitude to behave like $1/k^{2}$ asymptotically, as this 
is the leading order result if
one takes a one-gluon exchange interaction between the quark and antiquark.  
We introduce the spectral density function
$\rho(z)$, which takes a form different from that considered in 
Ref.\,\cite{Chang:2013pq}. We will see that different choices
of $\rho$ lead to different behaviors of
$q(x)$. 
The present algebraic model makes it possible for us 
to determine the $x$-dependence of the PDF. 

In Ref.\,\cite{Chang:2013pq}
it has been shown that $\rho(z)=\frac{1}{2}(\delta(1-z)+\delta(1+z))$ 
describes a bound state with point-particle-like
characteristics. 
It should be noted
that it also gives a constant PDF, if one calculates the PDF exactly, 
even though the Bethe-Salpeter amplitude is momentum dependent.
We infer that such behavior corresponds to the NJL prediction if one
performs a Pauli-Villas regularization. 
 
\begin{figure}[t]
\centerline{\includegraphics[width=0.8\linewidth]{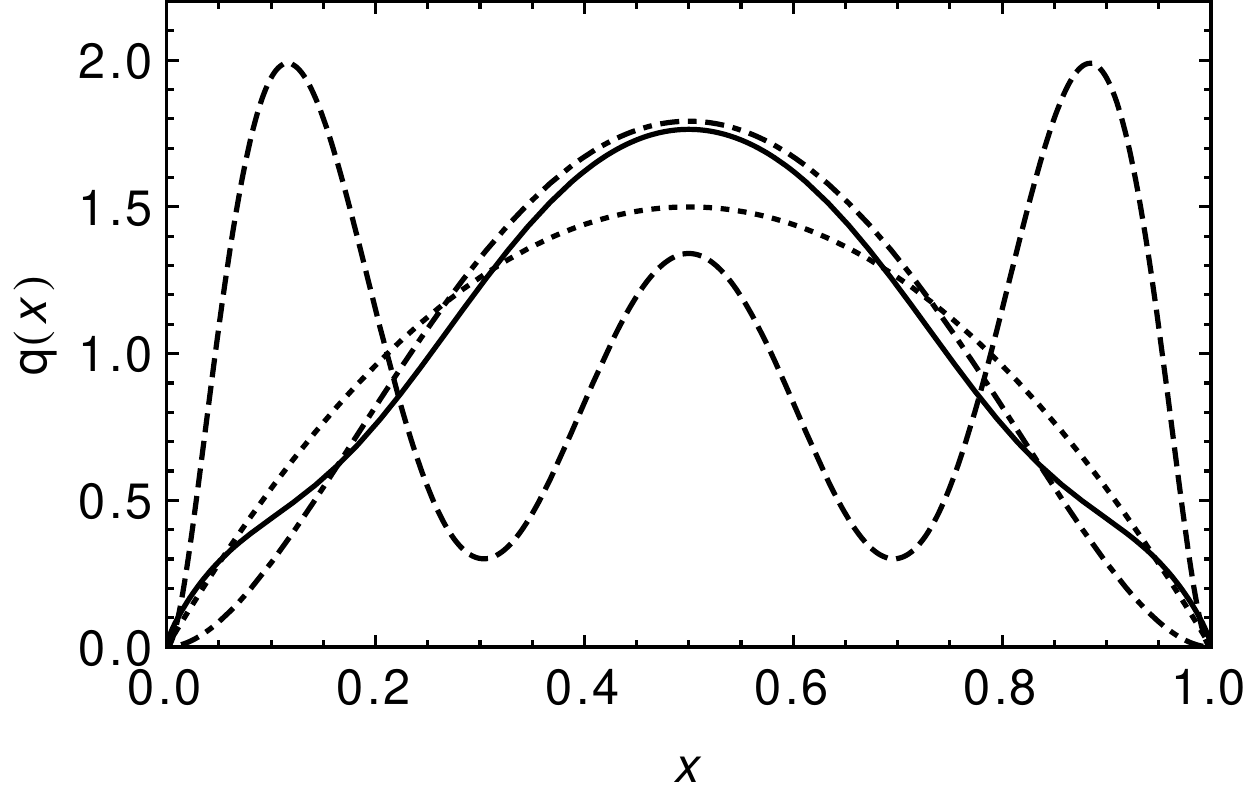}}
\caption{\label{fig:PDFmodel} Parton distribution function at $\zeta_{H}$. Curves:
\emph{solid},    Rainbow-ladder computation herein corresponding to Eq.(11);
\emph{dashed},\, $\rho(z)=\frac{3}{4}(1-z^{2})\left(1+6 a_{2}C_{2}^{3/2}(z)\right)$ with 
$a_2=0$ (\emph{dotdashed} $a_{2}=2/3$);
\emph{dotted}, \,$\rho(z)=\frac{1}{\pi}(1-z^{2})^{-\frac{1}{2}}$. }
\end{figure}

The QCD conformal limit can be 
reproduced with the spectral density $\rho(z)=\frac{3}{4}(1-z^{2})$. 
Ref.\,\cite{Chang:2014Basic} deduced a PDF which can
be approximately expressed by $30 x^{2} (1-x)^{2}$. Following 
this line, we extend the model of spectral
density as $\rho(z)=\frac{3}{4}(1-z^{2})\left(1+6 a_{2}C_{2}^{3/2}(z)\right)$,
where a second Gegenbauer polynomial
has been introduced and $a_{2}$ is a parameter. 
The corresponding PDA has the form 
$\varphi(x)=6 x(1-x)\left(1+a_{2}C_{2}^{3/2}(2x-1)\right)$. 
Obviously this form reproduces the
Chernyak-Zhitnitsky(CZ) form, with $a_{2}=2/3$\,\cite{cz:1984}. 
{}Following the method in Ref.\,\cite{Chang:2014Basic}, the PDF 
related to the CZ PDA can be 
computed consistently. The result is depicted 
in Fig.\,\ref{fig:PDFmodel}. Near $x=1$ this model has the power-law 
behavior, $(1-x)^{2}$, predicted by the QCD parton model. 
Here $a_{2}$ only affects the coefficient, not the $x$-dependence.
However, the PDF
shows oscillatory behavior that is difficult to reconcile with 
the physical meaning of the parton distribution function.
Of course, it is known that the CZ-like PDA is also double-humped 
and it can be argued that this is possible because it only 
has the interpretation of an amplitude.
In our consistent calculation we have show that the PDF is also  
double-humped and so we treat it with some caution.

In earlier work on the pion PDA we found that it is concave and broader 
than the asymptotic form at a typical hadronic scale.
To capture this characteristic we suggest another model for the 
spectral density, $\rho(z)=\frac{1}{\pi}(1-z^{2})^{-\frac{1}{2}}$,
which is
divergent
at $z=\pm 1$ but nevertheless integrable, $\int_{-1}^{1} \rho(z)=1$. 
The corresponding PDA and PDF are $\varphi(x)=\frac{8}{\pi}\sqrt{x(1-x)}$ 
and
$6 x (1-x)$, respectively.
This model for the PDA follows from
the precise mapping of string amplitudes in Anti-deSitter space 
to the light-front wavefunction of a hadron in 
physical space-time using holographic methods\,\cite{AdsQCD}. 
Such a PDA is also consistent numerical solution of the DSE,
although the actual power depends on the details of the interaction. 
It should be emphasised that the related PDF has the power-law
behavior, $1-x$, near $x=1$. That is in contrast to
the QCD parton prediction.  

Although the ultraviolet $k^2$ dependence is the same for the 
three different models described earlier, the parton distribution function
is very different. Ezawa~\cite{Ezawa:1974} predicted that the pion PDF 
would behave as $(1-x)^{2\alpha}$ if the pion
amplitude behaved as $\frac{1}{(k_{1}^{2})^{\alpha}}$, 
where $k_{1}$ is the struck quark momentum.

Let us write
a general form of the amplitude as 
$\int_{-1}^{1}dz (1-z^{2})^{\nu}\frac{1}{(k^{2}+z k\cdot P+M^{2})^{\beta}}$ 
with the relative
momentum $k$. If we fix one quark momentum and set the other to infinity 
we will find that the leading order amplitude is
$\frac{1}{(k_{1}^{2})^{1+\nu}}$ for $-1<\nu<0$, whatever the value 
of $\beta$. Based on Ezawa's work one readily finds
a PDF which behaves as $(1-x)^{2(1+\nu)}$, 
which is consistent with our results. 
If one works with a free gluon propagator that produces
an amplitude with $\nu=1; \beta=1$, then this yields the well known 
large-$x$ behavior, $(1-x)^{2}$. At the present time we can
only find a model independent formula if the relative momentum 
tends to infinity. However, the infrared interaction does effect 
the asymptotic form if one quark momentum goes to infinity with 
the other fixed. 

With the normalization constant set by hand, we can input the light 
quark mass at $\zeta=0.5$GeV and
arrange that it yields the correct pion mass. With an input current  
quark mass of $18.6$MeV we obtain $m_{\pi}=0.14$GeV and
$f_{\pi}=0.092$GeV. 
In the present work, the computation of the moments of the PDF 
is relatively straightforward because we employ 
algebraic parameterizations of the of quark propagator and 
the Bethe-Salpeter amplitude. The dressed-quark propagators are represented 
as\,\cite{Bhagwat:2003PDD}
\begin{equation}
S(p) = \sum_{j=1}^{j_m}\bigg[ \frac{z_j}{i \gamma\cdot p + m_j}+\frac{z^{\ast}}{i \gamma \cdot p + m_j^{\ast}}\bigg], \label{Spfit}
\end{equation}
with $\Im m_j \neq 0$ $\forall j$, so that $\sigma_{V,S}$ are 
meromorphic functions with no poles on the real $p^2$-axis,
a feature consistent with confinement \cite{Gribov:1999}. 
The pseudoscalar Bethe-Salpeter amplitude has the form
\begin{eqnarray}
\nonumber
\lefteqn{\Gamma(q;P) = \gamma_5
\big[ i E(q;P) + \gamma\cdot P F(q;P)   }\\
&&  \quad\quad  + \gamma\cdot q \, G(q;P) + \sigma_{\mu\nu} q_\mu P_\nu H(q;P) \big]. 
\label{BSK}
\end{eqnarray}
We retain all four terms in the pseudoscalar meson Bethe-Salpeter amplitude 
in the numerical calculation of the BS equation.
In the compuation of the PDF we find that the first two terms
in the BS amplitude dominate the parton structure. 
For this reason, for the main part of the present work we retain only
the first two terms and
leave the full calculation for future work.
We fit the associated scalar function via
{\allowdisplaybreaks
\begin{subequations}
\begin{eqnarray}
E(q;P) &=& E^{\rm i}(q;P) + E^{\rm u}(q;P) \,, \; \\
\nonumber E^{\rm i}(q;P) & = &
c^{\rm i}
\int_{-1}^1 \! dz \, \rho_{\nu^{\rm i}}(z) \bigg[
a \hat\Delta_{\Lambda^{\rm i}}^{4}(q_z^2) 
+ (1-a) \hat\Delta_{\Lambda^{\rm i}}^{5}(q_z^2)
\bigg], \label{Fifit}\\
\nonumber E^{\rm u}(q;P) & = & c^{\rm u} \int_{-1}^1 \! dz \, \rho_{\nu^{\rm u}}(z)
\hat\Delta_{\Lambda^{\rm u}}(q_z^2)\,, \quad\quad
\end{eqnarray}
\end{subequations}}
where $\rho_{\nu}(z) = \frac{\Gamma(\nu+\frac{3}{2})}{\sqrt{\pi}\Gamma(\nu+1)}(1-z^2)^{\nu}$
and $\Delta_{\Lambda}(q_z^2)=\frac{\Lambda^{2}}{q^{2}+z q\cdot P+\Lambda^{2}}$. This choice of 
denominator makes the fit easier when the meson mass is not zero.
The resulting parameter values are listed in Table~\ref{BSAparameters}.

\begin{table}[tb]
\centering
\caption{Fit parameters for the pseudoscalar meson Bethe-Salpeter amplitudes. \label{BSAparameters}}
\begin{tabular*}
{\hsize}
{
c|@{\extracolsep{0ptplus1fil}}
r@{\extracolsep{0ptplus1fil}}
r@{\extracolsep{0ptplus1fil}}
r@{\extracolsep{0ptplus1fil}}
c@{\extracolsep{0ptplus1fil}}
c@{\extracolsep{0ptplus1fil}}
l@{\extracolsep{0ptplus1fil}}
l@{\extracolsep{0ptplus1fil}}}
\hline
 & $c^{\rm i}$ & $c^{u}$ & $\phantom{-}\nu^{\rm i}$ & $\nu^{\rm u}$ & $a$\phantom{00} & $\Lambda^{\rm i}$ & $\Lambda^{\rm u}$ \\
\hline
$E_\pi$&5.8577&0.195&-0.656&1.08&2.682&1.247&1\\
$F_\pi$&2.9870&0.021&1.82&1.08&2.655&1.027&1\\
\hline
\end{tabular*}
\end{table}

In the actual calculations we use a small power to respect the possible 
anomalous dimension\,\cite{Chang:2013em} in the BS amplitude and the 
leading order behavior of the
amplitude in the ultraviolet 
region is $1/q^{2+2\alpha}$,
which is the natural outcome of using a one-gluon exchange interaction. 
One might conclude that this behavior will hold for any meson 
if the ultraviolet region is controlled by the one-gluon interaction. 
The leading-order spectral density, $\rho(z)$, in the ultraviolet 
region can be obtained exactly if the free gluon propgator is considered.
We choose the simple form to respect the infrared behavior.
The composition of power-4 and 5 in the
infrared part comes from the discovery of the behavior of 
the pion amplitude in the chiral limit. There the amplitude $E$ 
takes the same functional form as
the scalar part of the quark propagator, $B(q^{2})$. This composition 
provides a necessary condition for the existence of 
a point of inflection in $B(q^{2})$,
which is related to confinement\,\cite{inflexion}. 
For the light quark meson
the infrared power $\nu^{\rm i}<0$ in the spectral density,
yields an integrable singularity at $z=\pm 1$. 
We have shown that this is true for $\rho$ and $K$ mesons. 

Although we have been unable to prove that this choice of infrared behavior 
is unique, 
we favor it for the following reasons. 
Extending the UV analysis to the IR region suggests that there should 
be a singularity there~\cite{Farrar:PRL1979}.
On the other hand, in practice we usually fit $\nu^{\rm i}$ using 
the second Chebyshev moment of the amplitude. 
It should be noted that one can get the best fit by using a 
CZ-like spectral density with a positive value of $a_{2}$. We have
shown that this choice would produce an unlikely double humped 
parton distribution that is not well understood.

\begin{figure}[t]
\centerline{\includegraphics[width=0.8\linewidth]{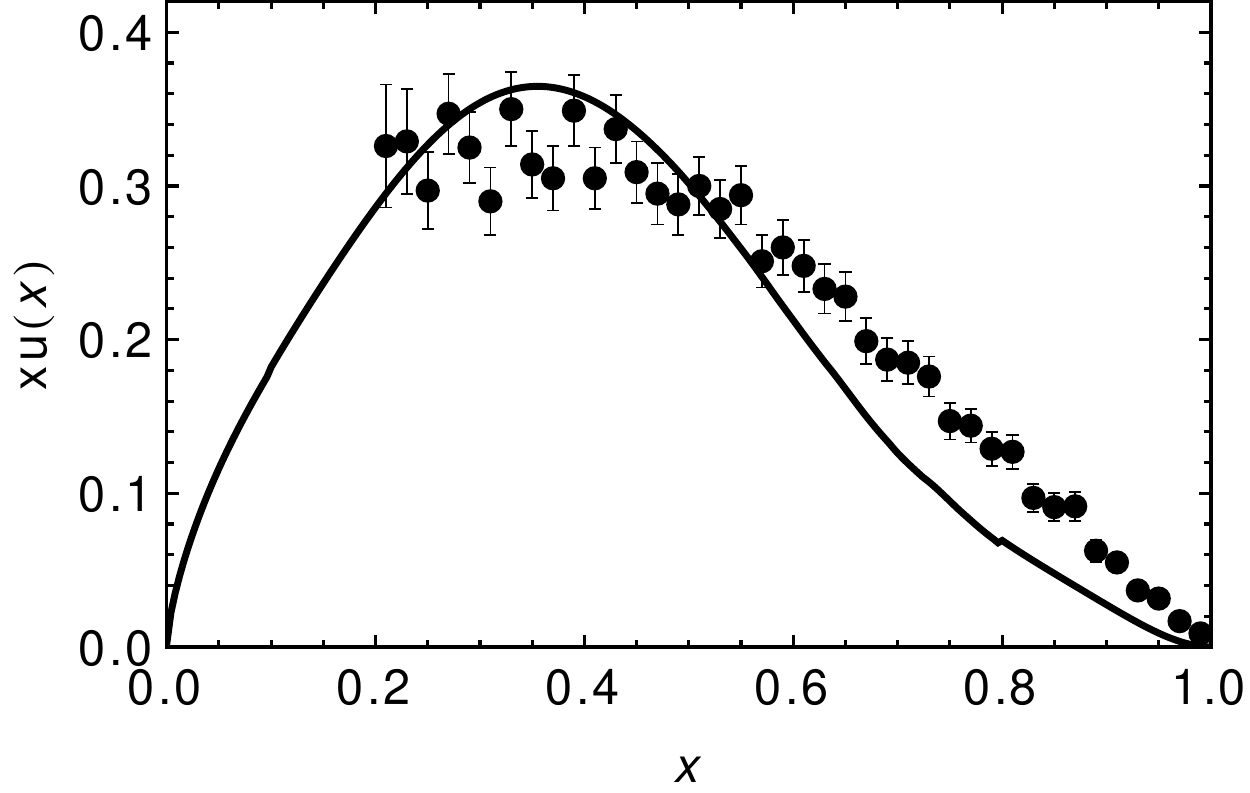}}
\caption{\label{fig:PDFrl} Comparision of the pion parton distribution function
calculated using the rainbow-ladder truncation computation described herein at $\zeta_{5}=5.2$GeV with
experimental data extracted from the $\pi$N Drell-Yan reaction\,\cite{Conway:1989}.}
\end{figure}

Based on the method we developed for calculating the PDA  
in Ref.\,\cite{Chang:2013pq}, it is straightforward to compute the PDF with
the quark propagator and meson amplitude in hand. 
The first step is to compute the moments 
$\langle x^m \rangle=\int_{0}^{1}dx x^{m} q(x)$. The algebraic form of
the input makes it possible to compute arbitrarily many moments. 
Considering the fact that PDF is an even function under 
$x\leftrightarrow(1-x)$ and vanishes at the endpoints unless 
the underlying interaction is momentum-independent, we
can reconstruct the PDF by expanding in  
Gegenbauer polynomials of order $\alpha$,  
that are a complete set with respect to the 
measure $(x(1-x))^{\alpha-1/2}$. Therefore, with complete generality, 
the PDF for $\pi$ 
may accurately be approximated as follows:
\begin{equation}
\label{PDFGalpha}
q(x) \approx  q_{m}(x) =N_\alpha [x \bar{x}]^{\alpha-1/2}
\bigg[ 1 + \sum_{j=2,4,\ldots}^{j_{\rm max}} a_j^\alpha C_j^\alpha(2 x - 1) \bigg],
\end{equation}
where $\bar{x}=1-x$, $N_\alpha = \Gamma(2\alpha+1)/[\Gamma(\alpha+1/2)]^2$. 
The parameters $\alpha, a_{j}$ can be fitted by the moments. 
In practice, this procedure converged very rapidly: $j_{m}=8$ was 
sufficient for the pion PDF. 
Our results for the PDFs are depicted in Fig.\,\ref{fig:PDFmodel} with 
the functions defined in Eqs.\,\eqref{PDFGalpha} and
\begin{equation}
\label{phiKB}
\begin{array}{lccccc}
   & \alpha  &  a_2 & a_{4} & a_{6} & a_{8}   \\
{\rm \pi} & 1.158 & -0.175 & 0.1 & -0.019 & -0.015 \\
\end{array}\,.
\end{equation}

The parton distribution function at $\zeta=0.5$GeV cannot be simply 
expressed by a one parameter representation like
$x^{\alpha}(1-x)^{\alpha}$. There is a point of inflection
around $x=0.85$, which can be thought of as the transition from
soft to hard scales. The PDF behaves as $(1-x)^{\nu}$ for $x>0.85$,
with $\nu\simeq 2 (1+\nu_{E}^{i})$,
consistent with our model analysis. The $F_{\pi}$ amplitude exhibits a 
positive spectral density power-law that would contribute
to the parton distribution as higher-twists. Including $F_{\pi}$ does not 
effect the region $x>0.85$ but does make the PDF more broad in the
infrared region.

In Fig.\,\ref{fig:PDFrl} we 
show the result of evolving the PDF, using the next-to-leading-order DGLAP 
equations\,\cite{BSR:Evolution}, from $\zeta_{H}\to \zeta_{5}$ with $\zeta_{5}=5.2$ GeV. 
The numerical results
favor a power-law in the valence region of the form $(1-x)^{1.6}$. 
At the initial scale we suppose the valence quarks carry all the pion momentum and 
have not attempted to include sea quark and gluon contributions
to provide a better description of the PDF in the soft region. That
could be done following the perspective suggested in Ref.\,\cite{Chang:2014Basic} 
without any difficulty. 

To summarize, we have presented the pion PDF within a 
rainbow-ladder truncation of the DSE approach. 
By employing a Nakanishi representation of Bethe-Salpeter amplitude 
and calculating the moments of the PDF to arbitrarily large values 
we have been able to calculate the $x$-dependence of the
PDF at a typical hadronic scale. We analyse 
the relation between the power-law behavior and the infrared interaction
that binds the meson. The present DSEs favor a power $(1-x)^{1.6}$,
at a typical experimental scale, $\zeta=5.2$GeV after
next-to-leading-order QCD evolution.

\medskip

\noindent\textbf{Acknowledgments}.
%
This work was supported by
the University of Adelaide and the Australian Research Council 
through an Australian Laureate Fellowship (AWT), FL0992247.
%




%





\end{document}